\documentclass[sort&compress,preprint,3p,number]{elsarticle}
\usepackage{amsmath,amssymb,xcolor}
\pdfoutput=1 % if your are submitting a pdflatex (i.e. if you have
\usepackage[pdfencoding=auto]{hyperref}
\usepackage[T1]{fontenc} % if needed
\usepackage{xspace}
\allowdisplaybreaks

\usepackage[normalem]{ulem}
\usepackage{multirow}
\usepackage{bashful}
\usepackage{todonotes}

\newcommand{\LiteRed}{\texttt{LiteRed}\xspace}
\newcommand{\Libra}{\texttt{Libra}\xspace}

\newcommand{\eeggg}{\ensuremath{e^+e^-\to 3\gamma}\xspace}
\newcommand{\eegg}{\ensuremath{e^+e^-\to 2\gamma}\xspace}
\newcommand{\eeph}{\ensuremath{e^+e^-\to 2\gamma(\gamma)}\xspace}

\newcommand{\minus}{\scalebox{0.75}[1.0]{-}}

\renewcommand{\S}{\mathcal{S}}

\newcommand{\e}{\ensuremath{\epsilon}}

\begin{document}
%\maketitle
%\flushbottom

\begin{frontmatter}
    
    \title{\boldmath Electron-positron annihilation to photons at $O(\alpha^3)$ revisited.}

    %% %simple case: 2 authors, same institution
    \author{Roman N. Lee}
    \ead{r.n.lee@inp.nsk.su}
    \address{Budker Institute of Nuclear Physics, 630090, Novosibirsk, Russia}

    \begin{abstract}
        We apply the modern multiloop methods to the calculation of the total cross sections of electron-positron annihilation to 2 and 3 photons exactly in $s/m^2$ with the accuracy $O(\alpha^3)$. Examining the asymptotics of our results, we find agreement with Ref. \cite{andreassi1962radiative} and discover mistakes in the results of Refs. \cite{eidelman1978e+,berends1981distributions}. This mistake is due to the terms, omitted in differential cross section in Refs. \cite{eidelman1978e+,berends1981distributions}, which are peaked in the kinematic region with all three photons being quasi-parallel to the collision axis. After restoring these terms, we find an agreement of the corrected result of Ref. \cite{berends1981distributions} with our result.
    \end{abstract}
    
\end{frontmatter}

\section{Introduction}
\label{sec:intro}
Modern methods of multiloop calculations crucially reduce the efforts required to check and improve the available results on radiative corrections. In this work we use this fortunate circumstance in order to calculate the total cross sections of the processes \eegg and \eeggg with accuracy $O(\alpha^3)$ for arbitrary energies. Surprisingly, we find that several results available in the ultrarelativistic limit contain errors. In particular, there is no correct result for the total cross section of \eeggg in the center-of-mass frame\footnote{Note that the frame dependence appears due to the restriction of the photon energies from below, necessary to avoid infrared divergence.}. Our technique is based on the Cutkosky rule which allows one to represent the phase-space integrals via the loop integrals with cut propagators. We apply the differential equations method to calculate the emerging two-loop integrals. We use the dimensional regularization $d=4-2\e$ to treat both infrared and ultraviolet divergences.

The paper is organized as follows. In the next section we present our results and discuss important issues related to them. Other sections contain details of the calculation. Th conclusion is presented in the last section.

\section{Results}\label{sec:results}

Let us present our results. Below we use the units $\hbar=c=m=1$, where  $m$ is the electron mass. Since the total cross sections $\sigma_{\eegg}$ and $\sigma_{\eeggg}$ are both infrared divergent, we define  $\sigma_{\eegg}(\omega_0)$ and $\sigma_{\eeggg}(\omega_0)$ which depend on the soft cut-off $\omega_0$. The quantity  $\sigma_{\eeggg}(\omega_0)$ is the cross section of the process \eeggg integrated over the kinematic region where the energy of any photon is greater than $\omega_0$.
The contribution of the complementary kinematic region (when the energy of one of the three photons is less than $\omega_0$) is then added to $\sigma_{\eegg}$ to form the finite quantity $\sigma_{\eegg}(\omega_0)$.
Note that the restriction of the integration region introduces the dependence of the cross section on the frame, which we denote by the upper superscript $\mathrm{f}$, as in $\sigma^{\mathrm{f}}_{\eeggg}(\omega_0)$.

In the center-of-mass frame we have
\begin{multline}\label{eq:CrossSection3cm}
    \sigma_{\eeggg}^{\mathrm{cmf}}(\omega_0)=
%3g/CrossSection3cm:
\frac{2\alpha}{\pi} \left(-\frac{1+\beta ^2}{2\beta} {\ln z}-1\right) {\ln  \left(\frac{\sqrt{s}}{2 \omega_0}\right)}\,\sigma _0(\beta)\\
    +\frac{\alpha ^3}{s \beta} \S\Bigg\{%\frac{\left(2 \beta +\left(\beta ^2+1\right) {\ln z}\right) \left(2 \beta  \left(\beta ^2-2\right)+\left(\beta ^4-3\right) {\ln z}\right)}{\beta ^2 }\ln \left(\frac{2 \omega_0}{\sqrt{s}}\right)
    \frac{4\left(3+\beta^4\right)}{s\beta^2}\bigg[
        4 \text{Li}_3(1-z)-2 \text{Li}_3(-z)-\left(2 \text{Li}_2(1-z)-\text{Li}_2(-z)\right){\ln z}-\frac32 \zeta_3%\text{Li}_3\left(z^2\right)-\text{Li}_2(z^2) {\ln z}+2\text{Li}_2(1-z) {\ln z}+\frac{1}{6}{\ln^3z}-\frac{\pi^2}{6}{\ln z}-\zeta_3
        \bigg]
       \\
    -\frac{16}{3s\beta} \left[{\ln^3z}+\pi ^2 {\ln z}\right]
    -\frac{4}{\beta}\left(s-2+\frac{16}{3 s}-\frac{8}{s^2}\right) \left[
        \text{Li}_2(-z)+ \frac{1}{2} {\ln s}\, {\ln z}
    \right]
    \\
    -\left(s\beta^2+\frac{7}{\beta^2}-\frac{2\beta^2+\beta^4}{3}\right){\ln^2z}
    +\frac{(s-4/s)\beta^2\pi ^2}{3}
    +\frac{8 (2+\beta^2)}{3 s\beta} {\ln z}
    +\frac{8 }{3s}
    \Bigg\}
%/3g/CrossSection3cm
.
\end{multline}
Here $\beta=\sqrt{1-4/s}$, $z=\frac{1-\beta}{1+\beta}$, 
\begin{equation}
\sigma_0(\beta)=\frac{\pi  \alpha ^2 }{s\beta}\left[-\frac{3-\beta ^4}{\beta}{\ln z}-2(2-\beta ^2)\right]
\end{equation}
is the Born cross section of the process \eegg, and we use the symmetrization symbol
\begin{equation}
    \S\left[ f(z,\beta)\right]\overset{\text{def}}{=}\frac12\left[f(z,\beta)+f(z^{-1},-\beta)\right]\,.
\end{equation}

It worth noting that $\beta \sigma_{\eeggg}^{\mathrm{cmf}}(\omega_0)$ is an analytical function of $\beta^2$ (or, equivalently, of $s-4$) in the vicinity of $\beta^2=0$.

The cross section of the process \eegg with the account of the first radiative correction has the form

\begin{multline}\label{eq:CrossSection2cm}
  \sigma_{\eegg}^{\mathrm{cmf}}(\omega_0)=
%2g/CrossSection2cm1:
  \left(1 +\frac{\pi \, \alpha}{v}\right)\sigma_0(\beta)
   -\frac{2\alpha}{\pi} \left(-\frac{1+\beta ^2}{2\beta} {\ln z}-1\right) {\ln  \left(\frac{\sqrt{s}}{2 \omega_0}\right)}\,\sigma _0(\beta)
   \\
+\frac{\alpha ^3}{s\beta}\S\Bigg\{
\frac{4 \left(2 s^2-s-9\right) }{s^2\beta}
\bigg[
-2\text{Li}_3\left(\frac{1}{1+z}\right)
-\frac12\text{Li}_2(-z){\ln s}
-\frac{1}{8} {\ln^2\!s}\, {\ln z}
-\frac16{\ln^3z}
-\frac{\pi^2}{8}{\ln z}
\bigg]
\\
+\frac{24 \left(s^2+s-3\right) }{s^2\beta}
\Re\bigg[
2\text{Li}_3\left(\frac{e^{\frac{i \pi }{3}}}{1+z}\right)
+\ln (s\,z)\text{Li}_2\left(\frac{e^{\frac{i \pi }{3}}}{1+z}\right)
-\frac14\text{Li}_2(-z){\ln s}
\\
-\frac18\left(
{\ln s}\,{\ln \frac{(s-1)^2}{s}}
+\frac{1}{6}{\ln^2z}
+\frac{\pi^2}{18}
\right){\ln z}
\bigg]
-\left(3-\beta^2\right)\left[
\text{Li}_2(-z) {\ln z}
+{\ln s}\, {\ln^2z}
\right]
\\
+\frac{2\left(1+\beta ^2\right) \left(3-\beta ^4\right)}{\beta^2}  \text{Li}_2(1-z){\ln z}
-\frac{1}{2\beta}\left[{\ln^2z}
-\pi^2\right] {\ln z}
\\
+\frac{9 s^2-8 s-4}{(s-1)s\beta}{\ln z}
+\frac{3 s-4}{s-1}{\ln s}
+\left(3-\beta^2\right)\frac{\pi^2}{4}
+\frac{16-5 \beta ^2-\beta ^4}{4 \beta ^2}{\ln^2z}
+\frac{20 }{s\beta}\text{Li}_2(-z)
\\
+\frac{4 (1+\beta^2) (2-\beta^2)}{\beta}\text{Li}_2(1-z)
+\left[\frac{3s+2}{s\beta }+\frac{3}{\beta  (s-1)^2}\right]{\ln s} {\ln z}
\Bigg\}
%/2g/CrossSection2cm1
.
\end{multline}
The first term here, $\left(1 +\frac{\pi \, \alpha}{v}\right)\sigma_0$, is nothing but the Born cross section $\sigma_0$, multiplied by the expansion of the Sommerfeld-Sakharov factor $\frac{2\pi \alpha/v}{1-e^{-2\pi \alpha/v}}$ with $v=\frac{2\beta}{1+\beta^2}$ being the relative velocity. It is remarkable that, apart from the contribution of term $\frac{\pi \, \alpha}{v}\sigma_0$, the cross section $\sigma_{\eegg}^{\mathrm{cmf}}(\omega_0)$, multiplied by $\beta$, is again an analytic function of $\beta^2$ in the vicinity of $\beta^2=0$.

The corresponding cross sections in the rest frame of the electron read

\begin{align}
    \sigma_{\eeggg}^{\mathrm{rf}}(\omega_0)=&\sigma_{\eeggg}^{\mathrm{cmf}}(\omega_0)+\delta\sigma,
    \label{eq:CrossSection3r}\\
     \sigma_{\eegg}^{\mathrm{rf}}(\omega_0)=&\sigma_{\eegg}^{\mathrm{cmf}}(\omega_0)-\delta\sigma.\label{eq:CrossSection2r}
\end{align}
where
\begin{multline}
\delta\sigma=
%3g/dCrossSection3r:
-\frac{\alpha ^3}{s\beta ^2}
\left[2 \beta  \left(\beta ^2-2\right)+\left(\beta ^4-3\right) {\ln z}\right]\S\Bigg\{
    \frac{1+\beta^2}{\beta }\left[
        \text{Li}_2(-z)%+\frac{{\ln^2z}}{4}
        +\frac{1}{2}{\ln s}\,{\ln z}% +\frac{\pi ^2}{12}
    \right]
    +\frac{2  {\ln z}}{s \beta}
    +1
\Bigg\}
%/3g/dCrossSection3r
.\label{eq:dsig}
\end{multline}

Note that the sum $\sigma_{\eeph}=\sigma_{\eeggg}+\sigma_{\eeggg}$  is independent of $\omega_0$ and, hence, of the frame.

\subsection{Asymptotics}
Let us now discuss the asymptotics of the presented results. 

\paragraph{Threshold asymptotics.} We start from the threshold asymptotics $\beta\ll 1$.

The threshold asymptotics of $\sigma_{\eeggg}^{\mathrm{cmf},\mathrm{rf}}(\omega_0)$ reads
\begin{equation}
   \sigma_{\eeggg}^{\mathrm{cmf},\mathrm{rf}}(\omega_0)=\frac{2\alpha^3}{3\beta}\left\{ \left(\pi ^2-9\right)+\left(-2 \ln\omega _0-\frac{31 \pi^2}{24}+8\right)\beta ^2 +O\left(\beta ^4\right)\right\}\,.
   \label{eq:eegggth}
\end{equation}
The first term in braces is well known and determines the orthopositronium decay width.

The threshold asymptotics of $\sigma_{\eegg}^{\mathrm{cmf},\mathrm{rf}}(\omega_0)$ reads
\begin{equation}
    \sigma_{\eegg}^{\mathrm{cmf},\mathrm{rf}}(\omega_0)-
    \left(1 +\frac{\pi \, \alpha}{v}\right)\sigma_0(\beta)=
    \frac{\alpha^3}{\beta}\left\{\frac{\pi ^2-20}{8}+ 
    \left(\frac{4}{3} \ln\omega_0-\frac{\pi
        ^2}{12}-\frac{20}{9}\right)\beta^2+O\left(\beta
    ^4\right)\right\}
    \label{eq:eeggth}
\end{equation}

The first term in braces is known for a long time, see, e.g., Ref. \cite{harris1957radiative}. In particular, this term determines the radiative correction to the parapositronium decay width. Note that the threshold expansion of $\delta\sigma$ in Eq. \eqref{eq:dsig} starts from $O(\beta^4)$, so Eqs. \eqref{eq:eegggth} and \eqref{eq:eeggth} hold both for center-of-mass frame and electron rest frame.

\paragraph{Ultrarelativistic limit.} Let us now discuss the high-energy asymptotics $s\gg 1$.

For the cross section $=\sigma_{\eeggg}$ we have
\begin{align}
    \sigma_{\eeggg}^{\mathrm{cmf}}(\omega_0)\approx&\frac{2 \alpha ^3 }{s}\left\{\left(2 {\ln    \frac{\sqrt{s}}{2 \omega_0}}-1\right)({\ln s}-1)^2+3-\frac{2 \pi^2}{3}\right\},\\
    \sigma_{\eeggg}^{\mathrm{rf}}(\omega_0)\approx&\frac{2 \alpha ^3 }{s}\left\{2 {\ln    \frac{\sqrt{s}}{2 \omega_0}}({\ln s}-1)^2+\frac{{\ln^3s}}{2}-\frac{3 {\ln^2s}}{2}-\frac{\pi ^2}{6}  \ln s+\ln s-\frac{\pi ^2}{2}+3\right\}\,.\label{eq:3ghigh}
\end{align}
The asymptotics of $\sigma_{\eeggg}^{\mathrm{rf}}(\omega_0)$ in the electron rest frame exactly coincides with the corresponding result of Ref. \cite{andreassi1962radiative}. However, the asymptotics of $\sigma_{\eeggg}^{\mathrm{cmf}}(\omega_0)$ in the center-of-mass frame does not coincide with the two available results \cite{eidelman1978e+,berends1981distributions}. Moreover, these two results differ from each other:
\begin{align}
    \sigma_{\eeggg}^{\mathrm{cmf}, \text{ Ref. \cite{eidelman1978e+}}}(\omega_0)\approx&\frac{2 \alpha ^3 }{s}\left\{\left(2 {\ln    \frac{\sqrt{s}}{2 \omega_0}}-1\right)({\ln s}-1)^2+3+\zeta_3\right\},\\
    \sigma_{\eeggg}^{\mathrm{cmf},\text{ Ref. \cite{berends1981distributions}}}(\omega_0)\approx&\frac{2 \alpha ^3 }{s}\left\{\left(2 {\ln    \frac{\sqrt{s}}{2 \omega_0}}-1\right)({\ln s}-1)^2+3\right\}\,.\label{eq:berends}
\end{align}
We have been able to trace the origin of discrepancy of our result with that of Ref. \cite{berends1981distributions}. Namely, it appeared that Refs. \cite{berends1981distributions,eidelman1978e+} have overlooked in the differential cross section the terms that contribute to the total cross section  in triply collinear kinematic region, see Appendix.% \ref{sec:UR}.

The ultrarelativistic asymptotics of $\sigma_{\eegg}$ reads
\begin{align}
    \sigma_{\eegg}^{\mathrm{cmf}}(\omega_0)-\sigma_0\approx&\frac{2 \alpha ^3 }{s}\left\{2 ({\ln s}-1)^2 {\ln \frac{2 \omega _0}{\sqrt{s}}}+\frac{{\ln ^3 s}}{6}+\frac{3 {\ln ^2s}}{4}+\left(\frac{\pi ^2}{3}-3\right) {\ln s}-\frac{\pi ^2}{12}\right\},\\
    \sigma_{\eegg}^{\mathrm{rf}}(\omega_0)-\sigma_0\approx
    &\frac{2\alpha ^3 }{s}\left\{2 ({\ln s}-1)^2 {\ln \frac{2 \omega _0}{\sqrt{s}}}-\frac{{\ln^3s}}{3} +\frac{5{\ln^2s} }{4}+\left(\frac{\pi ^2}{2}-2\right) {\ln s}-\frac{\pi ^2}{4}-1\right\}.
\end{align}
These two asymptotics coincide with the corresponding results of Refs. \cite{berends1981distributions} and \cite{andreassi1962radiative}, respectively.

\begin{figure}
    \centering\includegraphics[width=0.6\textwidth]{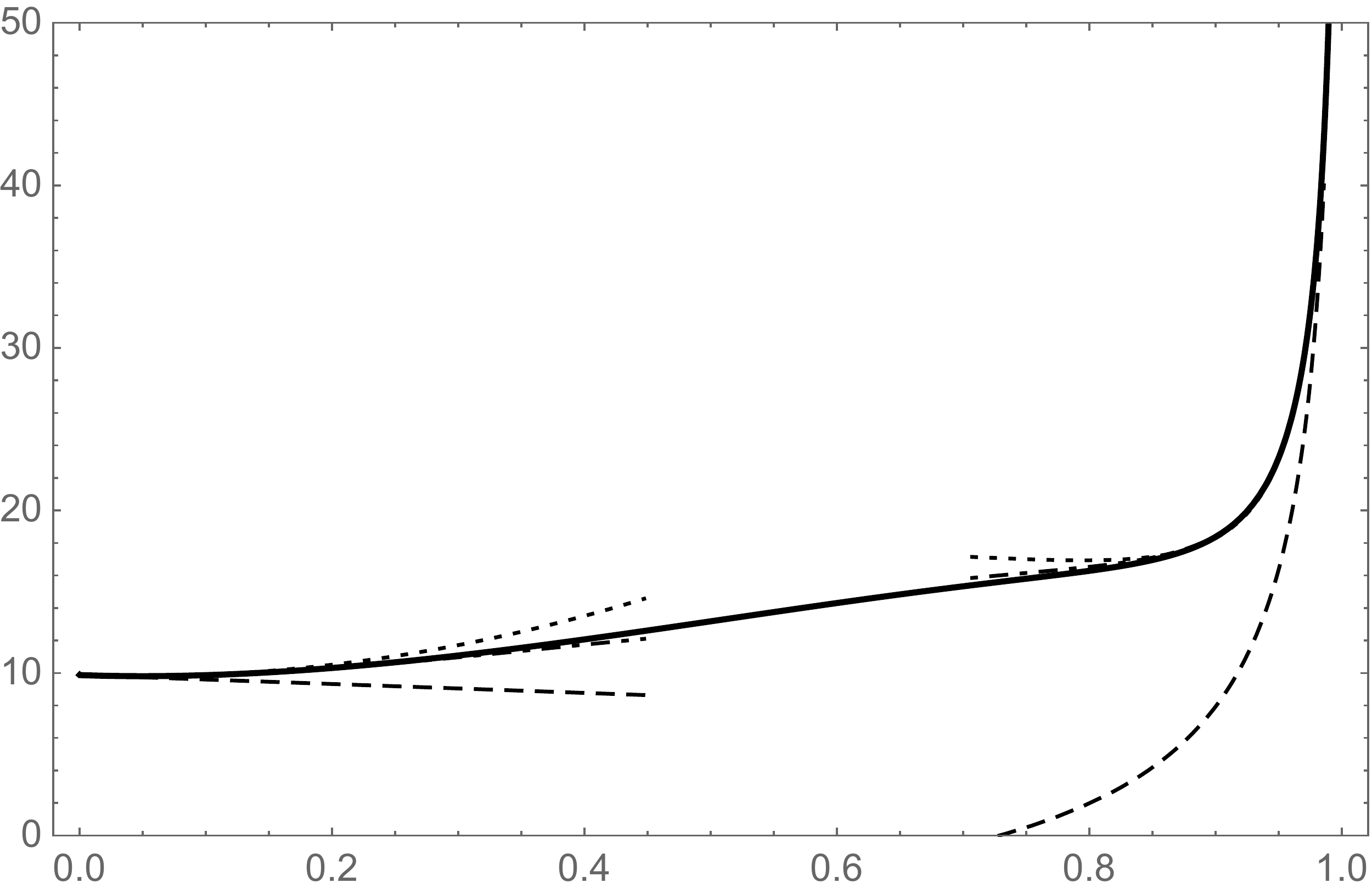}
    \caption{$O(\alpha^3)$ correction to $\sigma_{\eeph}$. Plotted: $(s-4m^2)[\sigma_{\eeph}]_{\alpha^3}$ as function of $\beta=\sqrt{1-4m^2/s}$. Dashed, dotted, and dash-dotted curves correspond to consecutive approximations of the threshold (truncation at $\beta^1,\beta^2,\beta^3$, left side of the graph) and high-energy (truncation at $(\frac1{s})^0,(\frac1{s})^1,(\frac1{s})^2$, right side of the graph) asymptotics.}
    \label{fig:CStot}
\end{figure}

The comparison of the exact cross section with the asymptotic expansions is demonstrated in Fig. \ref{fig:CStot}.
Let us present a few terms of high-energy expansion of the cross section $\sigma_{\eeph}$:
\begin{multline*}
    \sigma_{\eeph}=\frac{\alpha^3}{s-4}\bigg\{\frac{{\ln^3s}}{3}-\frac{{\ln^2s}}{2}+\frac{2}{3} \left(\pi ^2-3\right){\ln s}-\frac{3 \pi ^2}{2}+4\\
    +\frac{1}{s}\left(4 {\ln^2s}+\frac{26}{3}\left(\pi ^2-3\right){\ln s}+20 \zeta_3-\frac{22 \pi ^2}{9}-\frac{1}{3}\right)\\
    +\frac{1}{s^2}\left(9 {\ln^3s}+\frac{17}{2}{\ln^2s}-\frac{21\pi ^2+11}{3}{\ln s}+12 \zeta_3-\frac{41 \pi ^2}{6}+\frac{71}{18}\right)\\
    +\frac{1}{s^3}\left(-12 {\ln^3s}-\frac{137}{3} {\ln^2s}+\frac{156 \pi ^2-533}{9} {\ln s}+48 \zeta_3-\frac{266 \pi ^2}{9}+\frac{1961}{108}\right)+O(s^{-4})\bigg\}
\end{multline*}

It is remarkable that if we diminish by a factor of $2$ the term on the last line\footnote{This modification corresponds to taking a half-sum of two consecutive truncations, at $\left(\frac{1}{s}\right)^{2}$ and at  $\left(\frac{1}{s}\right)^{3}$.}, we will obtain an extremely good approximation for the exact cross section $\sigma_{\eeph}$ with the largest deviation about $2\%$ taking place at the threshold point. 

\section{\boldmath Calculation of \texorpdfstring{$\sigma_{\eeggg}$}{e+e- to 3 gamma cross section}.}\label{sec:sigma3}

\begin{figure}\centering
 \includegraphics[width=0.6\textwidth]{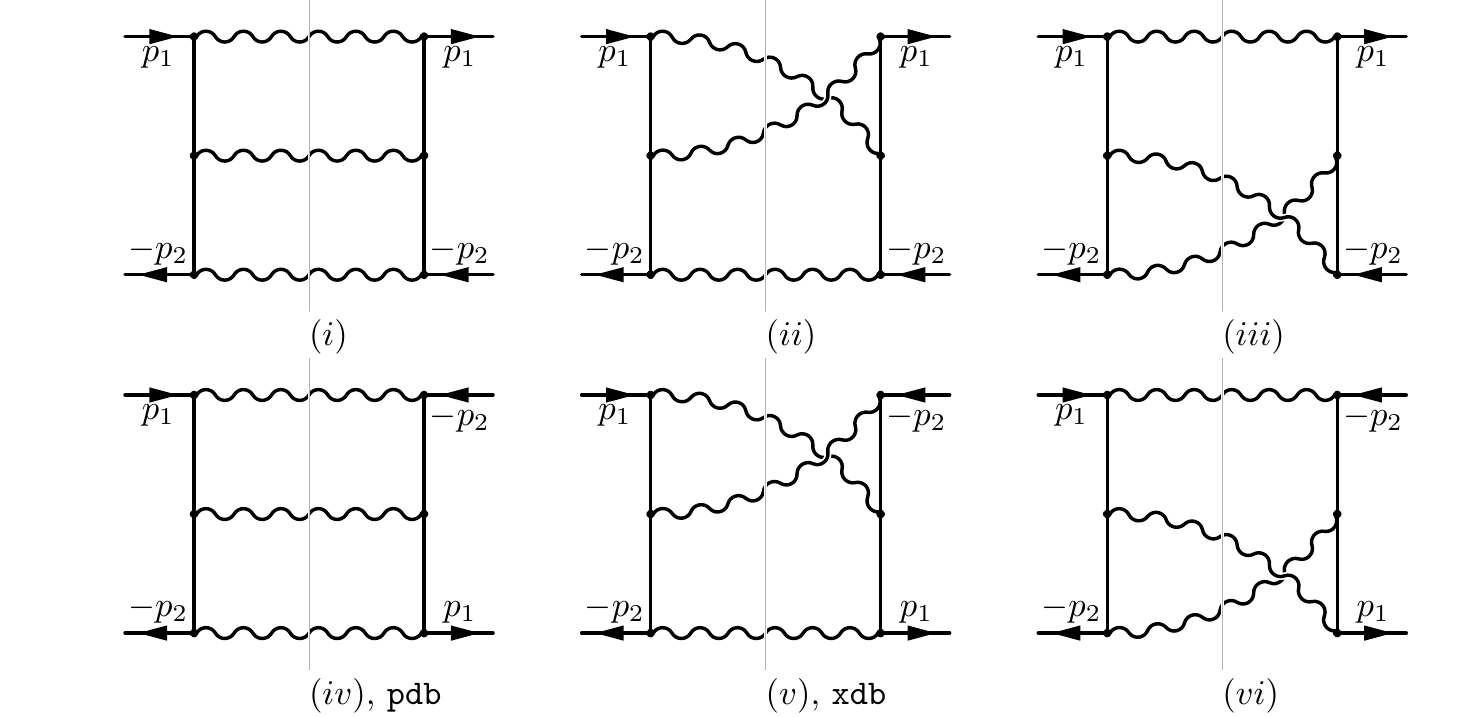}
 \caption{Diagrams contributing to the total Born cross section of \eeggg. Denominators of diagrams $(\romannumeral 4)$ and  $(\romannumeral 5)$ correspond to the \LiteRed{} bases \texttt{pdb} and \texttt{xdb}, respectively. \label{fig:amplitude3}}
\end{figure}

\begin{figure}
  \includegraphics[width=\textwidth]{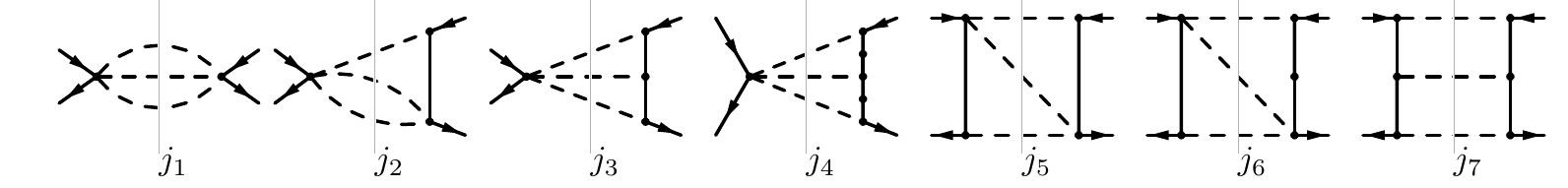}
  \caption{Master integrals enetring the cross section of \eeggg. Solid and dashed lines correspond to denominators $1-l^2$ and $-l^2$, respectively.\label{fig:mis3gamma}}
\end{figure}

We start with the calculation of the total Born cross section of the 3-photon annihilation\footnote{From now on we put the electron mass $m=1$ and recover the explicit dependence on it only in the final formulae on dimensional ground.}. The diagrams are shown in Fig. \ref{fig:amplitude3}.
We define two \LiteRed bases, \texttt{pdb} and \texttt{xdb}, corresponding to the denominators of diagrams $iv,\,v$ in Fig. \ref{fig:amplitude3}, respectively. These two bases are sufficient for the IBP reduction of all scalar integrals appearing in the cross section of the process \eeggg. There are 7 distinct master integrals which we choose as shown in Fig. \ref{fig:mis3gamma}.
We use \Libra\footnote{\texttt{Libra} package is available by request from the author.} package \cite{Libra} to reduce the system to $\e$-form \cite{Henn2013,Lee2014}. The new set of functions, $J_1,\ldots,J_7$ is related to $j_1,\ldots,j_7$ via
\begin{gather}
  J_1=\frac{4 j_1}{s}\,,\quad J_2=\frac{4 \beta  \e j_2}{2-3 \e } \,,\quad J_3=-\frac{2 \beta  s \e ^3j_3}{(1-2 \e ) (1-3 \e )_2} \,,\nonumber\\
  J_4=4  s \e \frac{(1+2 \e) j_4 - 2(1+\e ) \e ^2 j_3}{(1-2 \e ) (1-3 \e )_2}+\frac{4 \e (1+4 \e)j_2 }{2-3 \e } +\frac{12 (1+\e)j_1}{s} \,,\quad
  J_5=\frac{8 \e ^2\beta  s j_6}{(1-2 \e ) (1-3 \e )_2} \,,\nonumber\\
  J_6=-4\e ^2\frac{ s j_6 +(1-6 \e ) j_5}{(1-2 \e ) (1-3 \e
   )_2}-\frac{2 \e  j_2}{2-3 \e }\,,\quad J_7=\frac{2  \e ^3(s-4) s^2 j_7}{(1-2
   \e ) (1-3 \e )_2}\,,
\end{gather}
where $\beta=\sqrt{1-4/s}$ and $\alpha_n=\alpha\ldots (\alpha+n-1)$ is the Pochhammer symbol.
They satisfy the differential system in $\e$-form
\begin{equation}
  \partial_s \boldsymbol J = \e S \boldsymbol J\,,
\end{equation}
where
\begin{equation}
S={\left(
  \begin{array}{ccccccc}
    -\frac{2}{s} & 0 & 0 & 0 & 0 & 0 & 0 \\
    \frac{1}{\beta  s} & -\frac{s-8}{s^2\beta^2} & 0 & 0 & 0 & 0 & 0 \\
    \frac{2}{\beta  s} & 0 & \frac{2 (s-2)}{s^2\beta^2} & -\frac{1}{s\beta} & 0 & 0 & 0 \\
    0 & \frac{1}{s\beta} & \frac{4}{s\beta} & -\frac{2}{s} & 0 & 0 & 0 \\
    -\frac{1}{s\beta} & -\frac{3}{s} & 0 & 0 & -\frac{2 (s-6)}{s^2\beta^2} & -\frac{4}{s\beta} & 0 \\
    0 & 0 & 0 & 0 & -\frac{1}{s\beta} & -\frac{2}{s} & 0 \\
    \frac{4}{s\beta^2} & -\frac{3}{s\beta} & 0 & -\frac{2}{s\beta^2} & \frac{1}{s\beta} &  -\frac{2}{s\beta^2} & -\frac{2}{s} \\
  \end{array}
  \right)}\,.
\end{equation}
We fix the boundary conditions at $\beta\to 0$ ($s\to 4$) by evaluating the following coefficient in the asymptotics:
\begin{equation}
  \left[j_1\right]_{\beta^0}\,,\quad
  \left[j_4\right]_{\beta^0}\,,\quad
  \left[j_6\right]_{\beta^0}\,,\quad
  \left[j_2\right]_{\beta^{-1+2 \e}}=
  \left[j_3\right]_{\beta^{-1+2 \e}}=
  \left[j_5\right]_{\beta^{-1+2 \e}}=
  \left[j_7\right]_{\beta^{-2}}=0\,,
\end{equation}
where $\left[j_k\right]_{\beta^\mu}$ denotes to coefficient in front of $\beta^\mu$ in small-$\beta$ asymptotics of $j_k$ and we have explicitly indicated all coefficients which are obvious zeros. Thus, we are left with three nontrivial coefficients, $\left[j_{1,4,6}\right]_{\beta^0}$, which are nothing but the naive values of the corresponding integrals at the threshold, $j_{1,4,6}^{\text{th}}=j_{1,4,6}(s=4)$. Performing the IBP reduction we find that
\begin{equation}
  j_6^{\text{th}}=\frac{3 (1-3 \e)_2}{16 \e  (1+4 \e)} j_1^{\text{th}}+\frac{4 }{1+4 \e} j_4^{\text{th}}\,.
\end{equation}
The two remaining integrals $ j_{1,4}^{\text{th}}$ can be calculated exactly in $\e$ in terms of hypergeometric function, however we choose to follow the same approach as in Ref.\cite{lee2017dream} when calculating the parapositronium decay width to $4\gamma$. We choose the constant (but $\e$-dependent) overall normalization so that $j_1^{\text{th}}=1$. Then we have
\begin{align}
  % j_1^{\text{th}}&=\int \widetilde{dk_1} \widetilde{dk_2} (2\pi)^3\delta_+(k_1^2,k_2^2,k_3^2)
  % =\frac{\pi  2^{3-4 \e } \Gamma (1-\e )^3}{\Gamma (3-3 \e ) \Gamma (2-2 \e )},\\
  j_1^{\text{th}}&=1,
  \\
  j_4^{\text{th}}&=-\frac{1}{8 \e }+\frac{7}{16}+\frac{\pi ^2 \e }{48}+\frac{1}{96} \left(84 \zeta _3-54-\pi ^2\right) \e ^2+\frac{1}{96} \left(4 \pi ^4-42 \zeta _3-27 \pi ^2\right) \e ^3+O\left(\e ^4\right),
  \\
  j_6^{\text{th}}&=-\frac{1}{8 \e }+\frac{9}{16}+\left(\frac{\pi ^2}{12}-\frac{9}{16}\right) \e +\left(\frac{7 \zeta _3}{2}-\frac{3 \pi ^2}{8}\right) \e ^2+\left(\frac{\pi ^4}{6}-\frac{63 \zeta _3}{4}+\frac{3 \pi ^2}{8}\right) \e ^3+O\left(\e ^4\right).
\end{align}
This fully fixes the boundary conditions.
%It worth noting that $J_1,\ J_2,\ J_3,\ J_4,\ J_5,\ J_6,\ J_7$ starts from $O(\e^0),\ O(\e^1),\ O(\e^3),\ O(\e^0),\ O(\e^1),\ O(\e^2),\ O(\e^2)$, respectively.

Since the total cross section is infrared divergent at $\e=0$, we have to be careful with the  overall normalization, namely, we should pay attention to the factors which tend to unity as $\e\to 0$. We choose the following $n$-particle phase-space definition in $d=4-2\e$ dimensions:
\begin{equation}\label{eq:PS}
  d\Phi_n^{(4-2\e)} =\left(\frac{e^{\gamma_E}}{4\pi}\right)^{(n-1)\e}
  (2\pi)^{4-2\e}\delta^{(4-2\e)}\left(P_I-\sum\nolimits_{k=1}^{n}p_k\right)
  \prod_{k=1}^{n} \frac{d^{3-2\e} \boldsymbol{p}_k}{(2\pi)^{3-2\e}2\varepsilon_k}\,,
\end{equation}
where $\gamma_E=0.577\ldots$ is the Euler constant. The factor $\left(\frac{e^{\gamma_E}}{4\pi}\right)^{(n-1)\e}$ conveniently removes $\gamma_E$ and $\ln 4\pi$ in our intermediate formulae\footnote{
Since $n-1$ is the number of cut loops, we introduce $\left(\frac{e^{\gamma_E}}{4\pi}\right)^{\e}$ per each cut loop. This is exactly the factor which is introduced in the $\overline{\mathrm{MS}}$ scheme.}.
At $\e=0$ the definition turns into the usual definition of phase-space. We also normalize the trace of Dirac matrices by the condition
\begin{equation}
	\mathrm{Tr}\,1=4\,.
\end{equation}
%
%Then, in terms of $J_k$, the total cross section reads
%\begin{align}
%  \sigma=&\frac{\alpha ^3 2^{-4 \e } e^{2 \gamma  \e } \Gamma (1-\e )^3}{\beta  s \Gamma (3-3 \e ) \Gamma (2-2 \e )} \Bigg\{
%  \left(
%    -\frac{8 (s+2)}{s \e ^3}
%    +\frac{4 \left(51 s^3-62 s^2-392 s+256\right)}{3 (s-4) s^2 \e ^2}\right.\nonumber\\
%    &\left.
%    -\frac{8 \left(297 s^3-383 s^2-1376 s+1984\right)}{9 (s-4) s^2 \e }
%    +\frac{2 \left(10233 s^3-37336 s^2+51488 s-37120\right)}{27 (s-4) s^2}
%  \right)J_1\nonumber\\
%  &-2 \beta \left(
%    \frac{3 s^3-18 s^2-20 s+72}{3 (s-4) s \e ^2}
%    +\frac{-36 s^3+171 s^2+610 s-1452}{9 (s-4) s \e }
%  \right)J_2
%  -\frac{64 \beta  }{(s-4) \e ^3}J_3\nonumber\\
%  & +\left(\frac{4 (s+2)}{s \e ^3}
%  -\frac{2 \left(51 s^3-62 s^2-392 s+256\right)}{3 (s-4) s^2 \e ^2}+\frac{2 \left(585 s^3-766 s^2-2608 s+3968\right)}{9 (s-4) s^2 \e }\right.\nonumber\\
%  &\left.-\frac{8 \left(1242 s^3-4703 s^2+7174 s-4640\right)}{27 (s-4) s^2}\right)J_4
%  +2 \beta \left(\frac{36 s^3+99 s^2+344 s-516}{9 (s-4) s \e }\right.\nonumber\\
%  &\left.-\frac{3 s^3-6 s^2+16 s-24}{3 (s-4) s \e ^2}\right)J_5
%  +\left(\frac{4 (s+2)}{s \e ^3}-\frac{2 \left(6 s^4+3 s^3+10 s^2-200 s-128\right)}{3 (s-4) s^2 \e ^2}\right)J_6\nonumber\\
%  &+\left(\frac{4 \left(s^3-12 s+8\right)}{(s-4) s^2 \e ^3}-\frac{30 s^3-312 s+208}{(s-4) s^2 \e ^2}\right)J_7+O(\e)\Bigg\}\,.
%\end{align}
%Note that we have suppressed the terms which do not contribute in the limit $\e\to0$ in the coefficients.
Substituting the results for the master integrals $J_k$, we obtain
\begin{multline}\label{eq:cs3g}
    \sigma_{\eeggg}=
%3g/CrossSection: <-- this is needed for automatic check
  \frac{\alpha^3}{s}\Bigg\{\frac{1}{ \e}
  \frac{\left(2 \beta +\left(\beta ^2+1\right) {\ln z}\right) \left(2 \beta  \left(\beta ^2-2\right)+\left(\beta ^4-3\right) {\ln z}\right)}{2 \beta ^3 }
  \\
    -\frac{8  \beta  \left(s^2-2 s+4\right) }{(s-4)^2 s}\bigg(\text{Li}_3(z^2)-2\text{Li}_2(-z) {\ln z}+2 \ln (1-z) {\ln^2z}-2 {\ln s}\, {\ln^2z}
  \\
      -\frac{5}{6} {\ln^3z}-\frac{\pi^2}{2}{\ln z}-\zeta_3\bigg)
	+\frac{8 \beta  (s+2) }{s-4} {\ln z}\bigg(\text{Li}_2(1-z)+\frac{1}{4}{\ln^2z}+
   \frac{1}{2}{\ln s}\, {\ln z}\bigg)
  \\
    -\frac{16}{3 (s-4)} \left({\ln^3z}+\pi ^2  {\ln z}\right)
    +\frac{8  \left(s^2+3 s-8\right)}{(s-4) s}\left(\text{Li}_2(1-z)+\frac{1}{4}{\ln^2z}+{\ln s}\, {\ln z}\right)
  \\
    -4\left(s+\frac{17}{6 \beta ^2}-\frac{\beta ^2}{2}-\frac{1}{3}\right) \left( \text{Li}_2(-z)+\frac{1}{4}{\ln^2z}+\frac{1}{2}{\ln s}\, {\ln z}+\frac{\pi^2}{12}\right)
    +\frac{(s^2-4)\beta}{3 s}\pi ^2
  \\
    -\left(s+\frac{16}{(s-4)^2}+\frac{4}{3 s}\right)\beta {\ln^2z}
    +\frac{4 \left(3 s^2+21 s-8\right) {\ln z}}{3 (s-4) s}
    +\frac{4 \beta  (s+4)}{s-4}{\ln s}
    -\frac{2 \beta  (3 s-4)}{3  (s-4)}
  \Bigg\}\,,
%/3g/CrossSection <-- this is needed for automatic check
\end{multline}
where $z=\frac{1-\beta}{1+\beta}$.
Note that the cross section contains $\e^{-1}$ term, which is due to the infrared divergent contribution of the region where the energy of one of the outgoing photons is small. Thus, in order to obtain the finite quantity, we have to subtract the contribution of this region.  We derive the corresponding formulae in Section \ref{sec:soft}.

\section{Soft-photon contribution}\label{sec:soft}
The probability to emit soft photon is usually regulated by the fictitious photon mass. However, within our approach, we must stick to the dimensional regularization. As, to the best of our knowledge the relevant expressions are not in the literature, we derive them here with some details.

\subsection{Radiation probability.} We start from the following formula\footnote{The derivation of this formula is identical to that at $d=4$, see, e.g., Ref \cite{berestetskii1982quantum}.} for the probability of soft photon radiation:
\begin{equation}
  dW = -e^2\left(\sum_{n\in i\cup f} \sigma_n\frac{q_n p_n}{k\cdot p_n}\right)^2
  \left(\frac{e^{\gamma_E}}{4\pi}\right)^{\e}
  \frac{d^{3-2\e} \boldsymbol{k}}{(2\pi)^{3-2\e}2\omega}\,.
\end{equation}
Here $k=(\omega,\boldsymbol k)=(|\boldsymbol k|,\boldsymbol k)$ is the photon momentum, $\sum_{n\in i\cup f}$ denotes the sum over initial and final particles, $e_n=q_n|e|$ and $p_n=m_nu_n =m_n(\gamma_n,\gamma_n\boldsymbol{\beta}_n)$ are their charges and momenta (with $m_n=\sqrt{p_n^2}$ being the mass), and $\sigma_n=+1$ ($\sigma_n=-1$) when $n\in i$ ($n\in f$).
Note that we have again introduced a factor $\left(\frac{e^{\gamma_E}}{4\pi}\right)^{\e}$ for consistency with our previous definitions. The integration over $\omega$ will be restricted from above by the infrared cut parameter $\omega_0$ and can be trivially performed. Thus, we have
\begin{align}
W(\omega_0) =& \frac{2\alpha}{\pi}\underbrace{\intop_{0}^{\omega_0}\frac{2d\omega}{(2\omega)^{1+2\e}}}_{\left(2\omega_0\right)^{-2\e}/(-2\e)}\frac12\sum_{n,n'\in i\cup f}\left(-\sigma_n\sigma_{n'}\right)q_nq_{n'} I^{(4-2\e)}(u_n,u_{n'}),
\end{align}
where
\begin{equation}
  I^{(4-2\e)}(u_1,u_2) = e^{\e\gamma_E}\left(4\pi\right)^{-1+\e}\int  d\Omega \frac{\omega^2(u_1\cdot u_2)}{(k\cdot u_1)(k\cdot u_2)}= \frac{e^{\e\gamma_E}\Gamma(1-\e)}{\Gamma(2-2\e)}\int  \frac{d\Omega}{\Omega}\frac{\omega^2(u_1\cdot u_2)}{(k\cdot u_1)(k\cdot u_2)}\,.
\end{equation}
Here $\Omega=2\pi^{(d-1)/2}/\Gamma[(d-1)/2]$ and $\int \frac{d\Omega}{\Omega}\ldots$ denotes the averaging over the solid angle of $(3-2\e)$-dimensional vector $\boldsymbol k$. Once we put $\e=0$, it is easy to show that the integral is Lorentz invariant and evaluates to a well-known result (see, e.g., \cite{berestetskii1982quantum})
\begin{equation}
  I^{(4)} = \frac1{2\beta_{12}}\ln\frac{1+\beta_{12}}{1-\beta_{12}}\,,
\end{equation}
where $\beta_{12}=\sqrt{1-1/(u_1\cdot u_2)^2}$ is the relative velocity of the particles. However, we need also the $O(\e)$ term, and this term is frame-dependent. In general, the integral depends on three parameters, $\beta_{12}$, $\beta_1=|\boldsymbol{u}_1|/u_1^0$, and $\beta_2=|\boldsymbol{u}_2|/u_2^0$ ($\beta_{1,2} $ are the velocities of the particles in the lab frame).
Remarkably, it is possible to calculate the \e-expansion of $I^{(4-2\e)}$ using the multiloop methods. Consider the family of integrals
\begin{equation}
	\widehat{j}(n_1,n_2,n_3,n_4) =
	\int \frac{2 d^d k}{\Omega(k\cdot u_1)^{n_1}(k\cdot u_2)^{n_2}}
	\frac 1\pi\Im\left[(1-k\cdot u_0-i0)^{-n_3}\right]
	\frac 1\pi\Im\left[(-k^2-i0)^{-n_4}\right],
\end{equation}
where $u_0=(1,\boldsymbol{0})$ and we assume that $(u_{1,2}\cdot u_0)>0$.
It is easy to see that
\begin{equation}
  I^{(d)} = \frac{e^{\e\gamma_E}\Gamma(1-\e)}{\Gamma(2-2\e)}(u_1\cdot u_2)\widehat{j}(1,1,1,1).
\end{equation}
Performing the IBP reduction, we find four master integrals. We pass to variables
$x_1=\sqrt{\frac{1-\beta_1}{1+\beta_1}}$, $x_2=\sqrt{\frac{1-\beta_2}{1+\beta_2}}$, $x_3=\sqrt{\frac{1-\beta_{12}}{1+\beta_{12}}}$,
and reduce the differential systems with respect to $x_1$, $x_2$, and $x_3$ to $\e$-form (using \Libra) and find the following `canonical' basis:
\begin{equation}
	\widehat{J}_1={\scriptstyle \widehat{j}(0,0,1,1)},\ \widehat{J}_2=\frac{\e(1-x_2^2)}{(1-2\e)x_2}{\scriptstyle \widehat{j}(0,1,1,1)},\ \widehat{J}_3=\frac{\e(1-x_1^2)}{(1-2\e)x_1}{\scriptstyle \widehat{j}(1,0,1,1)},\ \widehat{J}_4=\frac{\e(1-x_3^2)}{(1-2\e)x_3}{\scriptstyle \widehat{j}(1,1,1,1)}.
\end{equation}
They satisfy the following differential system in Pfaff form
\begin{gather}
	d\widehat{\boldsymbol{J}} = \e\,d\left(
	\begin{array}{cccc}
        0 & 0 & 0 & 0 \\
        -2\ln{x_2} & 2\ln{\frac{1-x_2^2}{x_2}} & 0 & 0 \\
        -2\ln{x_1} & 0 & 2\ln{\frac{1-x_1^2}{x_1}} & 0 \\
        -2\ln{x_3} &
        \ln{\frac{\tilde{x}_1\tilde{1}}{\tilde{x}_3\tilde{x}_2}} &
        \ln{\frac{\tilde{x}_2\tilde{1}}{\tilde{x}_3\tilde{x}_1}} &
        \ln{\frac{\tilde{x}_3\tilde{x}_2\tilde{x}_1\tilde{1}}{x_1^2 x_2^2 \left(1-x_3^2\right)^2}}
	\end{array}
	\right)\widehat{\boldsymbol{J}},
\end{gather}
where we have used the notation $\tilde{a}=a-x_1x_2x_3/a$. The physical region is defined by the inequalities
\[0\leqslant x_\rho\leqslant 1,\quad \tilde{x}_\rho\geqslant0\,.\]
We fix the boundary conditions at the point $x_1=x_2=x_3=1$ and travel to the generic point $(x_1,x_2,x_3)$ in the physical region along the contour $\gamma(0\leqslant \tau\leqslant 2$) defined piece-wise as
\begin{equation}
	\gamma(\tau)=\begin{cases}
	\left(1-\tau+\tau x_1,1-\tau+\tau x_2,(1-\tau+\tau x_1)(1-\tau+\tau x_2)\right)\,,& 0\leqslant \tau\leqslant 1\\
	\left(x_1, x_2,x_1 x_2+(\tau-1)(x_3-x_1 x_2)\right)\,,& 1< \tau\leqslant 2\\
	\end{cases}
\end{equation}
The boundary conditions appear to be trivial with the only nonzero constant being \[\widehat{J}_1\bigg|_{x_1=x_2=x_3=1}=1\,.\]
We finally obtain
%\todo[color=yellow!40]{manually checked}
\begin{align}
	\widehat{J}_1&=1,\quad  \widehat{J}_2 (x)=\widehat{J}_3 (x)=J_4 (x,1,x),\nonumber\\
	\widehat{J}_4 &(x_1,x_2,x_3) = -2 \e {\ln x_3} \nonumber\\
	   &+2\e^2\left[
       f\left({x_1 x_3}/{x_2}\right)+f\left({x_2 x_3}/{x_1}\right)+f\left(x_1 x_2 x_3\right)-f\left({x_1 x_2}/{x_3}\right)-f\left(x_3^2\right)
		\right]
         +O\left(\e^3\right)\,,
\end{align}
where
\begin{equation}
    f(x)=\text{Li}_2(1-x)+\frac{1}{4}{\ln^2x}\,.
\end{equation}
Finally, we obtain
\begin{align}
I^{(4-2\e)}(u_1,u_2) =&F(x_1,x_2,x_3)=\frac1{\beta_{12}}\bigg\{-{\ln x_3}
+\e\bigg[
f\left({x_1 x_3}/{x_2}\right)+f\left({x_2 x_3}/{x_1}\right)+f\left(x_1 x_2 x_3\right)\nonumber\\
&\hspace{6cm}-f\left({x_1 x_2}/{x_3}\right)-f\left(x_3^2\right)\bigg]\bigg\}+O\left(\e^2\right)\,,\nonumber\\
I^{(4-2\e)}(u_1,u_1) =&F(x_1,x_1,1)=1-\epsilon \frac{2}{\beta_1} \ln x_1+O\left(\e^2\right)\,,
\end{align}
where
\begin{gather}
	x_1=u_1^0-|\boldsymbol{u}_1|=\gamma_1(1-\beta_1), \quad
	x_2=u_2^0-|\boldsymbol{u}_2|=\gamma_2(1-\beta_2),\nonumber\\
	x_3=u_1\cdot u_2-\sqrt{(u_1\cdot u_2)^2-1}=\gamma_{12}(1-\beta_{12}),\quad \gamma_\rho=1/\sqrt{1-\beta_\rho^2}.
\end{gather}
\subsection{Soft-photon contribution to \texorpdfstring{\eeggg}{e+e- to 3 gamma}.}
Let us now derive the cross section of \eeggg integrated over the kinematic region where the energies of all photons are restricted from below by some experimental cut-off $\omega_0$. This restriction obviously introduces the frame dependence, and we will specialize our formulae to two physically relevant frames: the center-of-mass frame and the rest frame of the initial electron.

The cross section $\sigma_{\eeggg}(\omega_i>\omega_0)$ is obtained by subtracting from $\sigma_{\eeggg}$ the contribution of the soft region:
\begin{equation}
    \sigma_{\eeggg}^{\mathrm{f}}(\omega_0)=\sigma_{\eeggg}-W^{\mathrm{f}}(\omega_0)\sigma_0\,,
\end{equation}
where $\mathrm{f}=\mathrm{cmf}$ and $\mathrm{f}=\mathrm{rf}$ for the center-of-mass frame and the electron rest frame, respectively.
We have
\begin{align}
	W^{\mathrm{cmf}}(\omega_0)=&\frac{2\alpha}{\pi}\frac{\left(2\omega_0\right)^{-2\e}}{(-2\e)}\left[
	F(\sqrt{z},\sqrt{z},z)-F(\sqrt{z},\sqrt{z},1)
	\right]\,,\nonumber\\
%	=&\frac{2\alpha}{\pi}\frac{\left(2\omega_0\right)^{-2\e}}{(-2\e)}\left[
%    -\frac{\beta  (s-2) {\ln z}}{s-4}-1
%    +\epsilon  \left(\frac{\beta  (s-2) }{2(s-4)}\left[4\text{Li}_2(1-z)+{\ln^2z}\right]+\frac{\beta  s {\ln z}}{s-4}\right)\right]\,,\\
	W^{\mathrm{rf}}(\omega_0)=&\frac{2\alpha}{\pi}\frac{\left(2\omega_0\right)^{-2\e}}{(-2\e)}\left[
	F(z,1,z)-\tfrac12F(z,z,1)-\tfrac12F(1,1,1)
	\right].\label{eq:Wf}
%	=&\frac{2\alpha}{\pi}\frac{\left(2\omega_0\right)^{-2\e}}{(-2\e)}\left[-\frac{\beta  (s-2) {\ln z}}{s-4}-1+\epsilon  \left(\frac{\beta  (s-2)}{s-4}
%	\left[\text{Li}_2\left(1-z^2\right)+{\ln^2z}+{\ln z}\right]
%	-1\right)\right]\,.
\end{align}
The two-photon annihilation Born cross section $\sigma_{\eegg}$ should also be calculated with $\e^1$ terms retained:
\begin{multline}
	\sigma(\eegg)=
%2gBorn/CrossSection: <--- for automatic check of the formulae
    \frac{\pi\, \alpha ^2}{s} \Bigg\{\frac{2 \beta  \left(\beta ^2-2\right)+\left(\beta ^4-3\right) {\ln z}}{\beta ^2}\\
	+2\epsilon\,  \bigg(
	\frac{3-\beta^4}{\beta^2}\left[\text{Li}_2(1-z)+\frac14{\ln^2 z}+\frac{1}{2}{\ln s}\, {\ln z}\right]
    +\frac{3-\beta^2}{\beta^2}{\ln z}
	+\frac{2-\beta^2}{\beta}{\ln s}
    -\frac{1}{\beta}
    \bigg)\Bigg\}
%/2gBorn/CrossSection <--- for automatic check of the formulae
.
\end{multline}
We finally arrive at Eq. \eqref{eq:CrossSection3cm}.

\section{\boldmath Calculation of \texorpdfstring{$\sigma_{\eegg}$}{e+e- to 2 gamma cross section} at NLO.}\label{sec:sigma2}

\begin{figure}
  \centering
  \includegraphics[width=\textwidth]{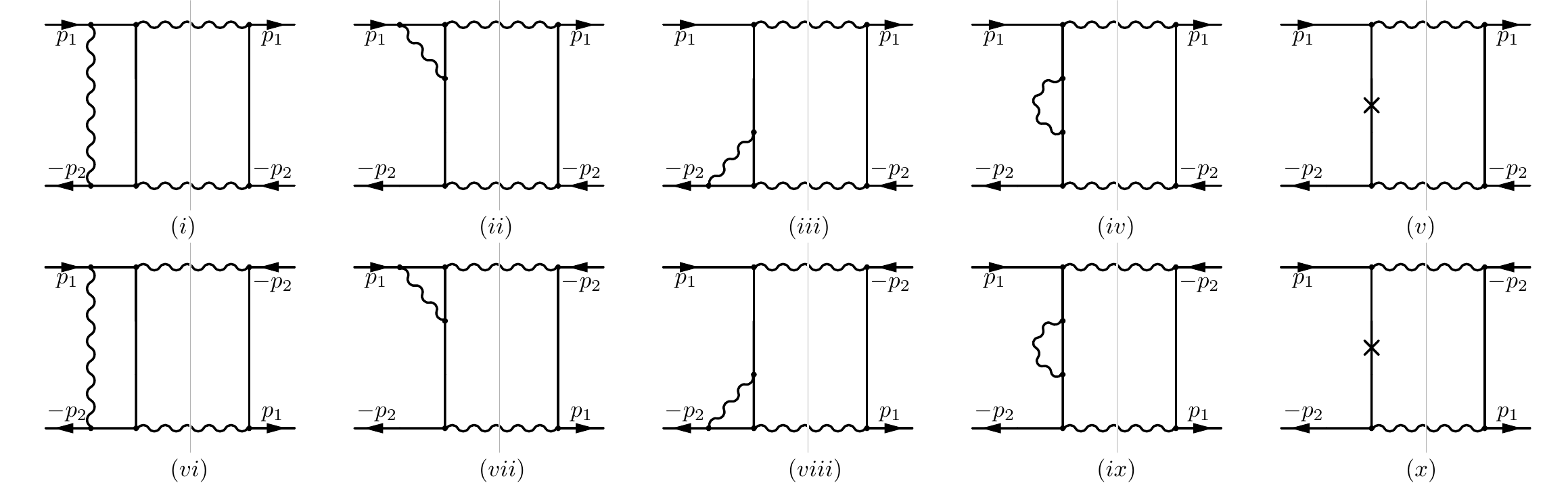}
   \caption{Diagrams contributing to the virtual correction to the total cross section of \eegg. Denominators of diagrams $(\romannumeral 1)$ and  $(\romannumeral 6)$ correspond to the \LiteRed bases \texttt{pdb2} and \texttt{pdb1}, respectively. The diagrams $(\romannumeral 5)$ and  $(\romannumeral 10)$ correspond to the mass counterterm.
   \label{fig:amplitude2}}
\end{figure}

\begin{figure}
  \centering
  \includegraphics[width=\textwidth]{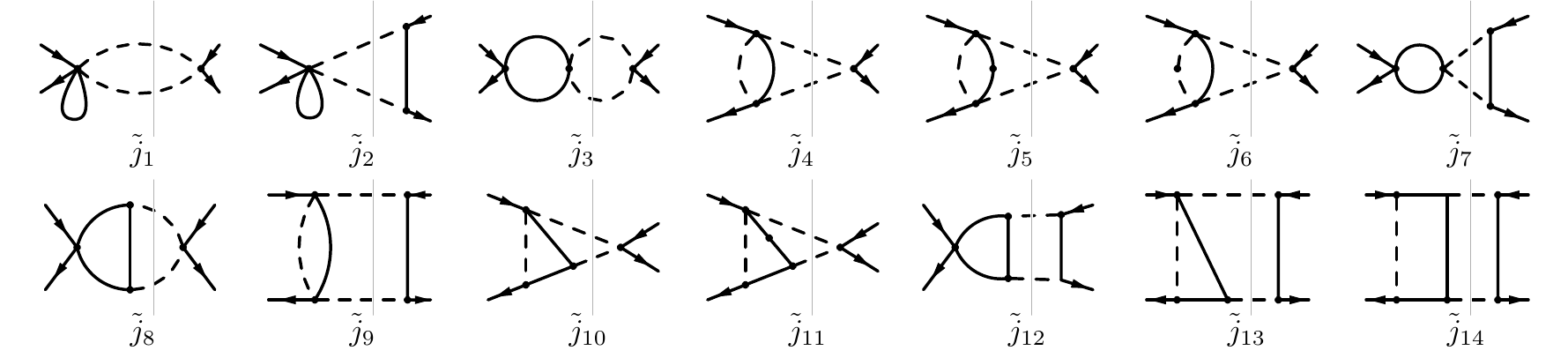}
   \caption{Two-loop (one loop cut) master integrals for the virtual correction to the total cross section of \eegg.
   \label{fig:mis2gamma}}
\end{figure}

Let us now briefly describe the calculation of the virtual correction to the total cross section of $\eegg$.
We calculate the contribution of the diagrams depicted in Fig. \ref{fig:amplitude2}.
The IBP reduction of the two-loop diagrams reveals 14 master integrals depicted in Fig.  \ref{fig:mis2gamma}.
We use \Libra to reduce the differential system for $\tilde{j}_{1-14}$ to $\e$-form.
The `canonical' master integrals $\tilde{J}_{1-14}$ are defined as follows
\begin{gather}
\tilde{J}_1=\tilde{j}_1,\quad
\tilde{J}_2=-\frac{\beta  s \e  \tilde{j}_2}{2 \e -1},\quad
\tilde{J}_3=\frac{(1-2 \e) \tilde{j}_3}{\beta  (1-\e)}+\frac{\tilde{j}_1}{\beta },\quad
\tilde{J}_4=-\frac{s\beta \e  \tilde{j}_6}{2 \left( 1-2 \e \right) _2},
\nonumber\\
\tilde{J}_5=\frac{s\beta \e  \left(\tilde{j}_6+2\tilde{j}_5\right)}{2\left( 1-2 \e \right) _2},\quad
\tilde{J}_6=\frac{s \e  \tilde{j}_5-2 (1-4 \e) \tilde{j}_6}{\left( 1-2 \e \right) _2}-\frac{(1-3 \e) \tilde{j}_4}{1-\e},\quad
\tilde{J}_7=\frac{s \e  \tilde{j}_7+4 \e  \tilde{j}_1}{1-\e}+\frac{s \e  \tilde{j}_2}{1-2 \e},
\nonumber\\
\tilde{J}_8=\frac{s \e  \tilde{j}_8-4 \e  \tilde{j}_1}{1-\e},\quad
\tilde{J}_9=\frac{\beta  (s-1) \e  \tilde{j}_9}{1-\e}-\frac{s\beta^3\e  \tilde{j}_6}{2 \left( 1-2 \e \right) _2}-\frac{\beta  \e  \tilde{j}_2}{1-2 \e},\quad
\tilde{J}_{10}=\frac{s \tilde{j}_{11}-2 s \e ^2 \tilde{j}_{10}+s \e  \tilde{j}_5}{\left( 1-2 \e \right) _2},
\nonumber\\
\tilde{J}_{11}=\frac{2 s\beta \e ^2 \tilde{j}_{10}}{ \left( 1-2 \e \right) _2},\quad
\tilde{J}_{12}=\frac{2 s^2\beta \e ^2 \tilde{j}_{12}}{ \left( 1-2 \e \right) _2},\quad
\tilde{J}_{13}=-\frac{2s^2\beta \e ^2 \tilde{j}_{13}}{\left( 1-2 \e \right) _2},\quad
\tilde{J}_{14}=\frac{(s-4) s^2 \e ^2 \tilde{j}_{14}}{\left( 1-2 \e \right) _2}
\end{gather}
They satisfy the differential system
\begin{equation}
  \partial_{\beta} \tilde{\boldsymbol J} =  \e\left[\frac{\tilde{S}_0}{\beta}+\frac{\tilde{S}_1}{1-\beta^2} +\frac{\beta\tilde{S}_2}{1-\beta^2} +\frac{\beta\tilde{S}_3}{3+\beta^2} \right]\tilde{\boldsymbol J}\,,
\end{equation}
 where
 \begin{gather}
   \scriptsize
   \arraycolsep=1pt\def\arraystretch{0.3}
   \tilde{S}_0=\left[
   \begin{array}{cccccccccccccc}
    0 & 0 & 0 & 0 & 0 & 0 & 0 & 0 & 0 & 0 & 0 & 0 & 0 & 0 \\
    0 & 2 & 0 & 0 & 0 & 0 & 0 & 0 & 0 & 0 & 0 & 0 & 0 & 0 \\
    0 & 0 & \minus2 & 0 & 0 & 0 & 0 & 0 & 0 & 0 & 0 & 0 & 0 & 0 \\
    0 & 0 & 0 & 2 & 0 & 0 & 0 & 0 & 0 & 0 & 0 & 0 & 0 & 0 \\
    0 & 0 & 0 & 0 & 2 & 0 & 0 & 0 & 0 & 0 & 0 & 0 & 0 & 0 \\
    0 & 0 & 0 & 0 & 0 & 0 & 0 & 0 & 0 & 0 & 0 & 0 & 0 & 0 \\
    0 & 0 & 0 & 0 & 0 & 0 & 0 & 0 & 0 & 0 & 0 & 0 & 0 & 0 \\
    0 & 0 & 0 & 0 & 0 & 0 & 0 & 0 & 0 & 0 & 0 & 0 & 0 & 0 \\
    0 & 0 & 0 & 0 & 0 & 0 & 0 & 0 & 2 & 0 & 0 & 0 & 0 & 0 \\
    0 & 0 & 0 & 0 & 0 & 0 & 0 & 0 & 0 & 0 & 0 & 0 & 0 & 0 \\
    0 & 0 & 0 & 0 & 0 & 0 & 0 & 0 & 0 & 0 & 2 & 0 & 0 & 0 \\
    0 & 0 & 0 & 0 & 0 & 0 & 0 & 0 & 0 & 0 & 0 & 2 & 0 & 0 \\
    0 & 0 & 0 & 0 & 0 & 0 & 0 & 0 & 0 & 0 & 0 & 0 & 2 & 0 \\
    0 & 0 & 0 & 0 & 0 & 0 & 0 & 0 & 0 & 0 & 0 & 0 & 0 & 0 \\
   \end{array}
   \right],\
   \tilde{S}_1=\left[
\begin{array}{cccccccccccccc}
 0 & 0 & 0 & 0 & 0 & 0 & 0 & 0 & 0 & 0 & 0 & 0 & 0 & 0 \\
 2 & 0 & 0 & 0 & 0 & 0 & 0 & 0 & 0 & 0 & 0 & 0 & 0 & 0 \\
 2 & 0 & 0 & 0 & 0 & 0 & 0 & 0 & 0 & 0 & 0 & 0 & 0 & 0 \\
 1 & 0 & 0 & 0 & 0 & \minus1 & 0 & 0 & 0 & 0 & 0 & 0 & 0 & 0 \\
 0 & 0 & 0 & 0 & 0 & 1 & 0 & 0 & 0 & 0 & 0 & 0 & 0 & 0 \\
 0 & 0 & 0 & 6 & 2 & 0 & 0 & 0 & 0 & 0 & 0 & 0 & 0 & 0 \\
 0 & 2 & 2 & 0 & 0 & 0 & 0 & 0 & 0 & 0 & 0 & 0 & 0 & 0 \\
 0 & 0 & 2 & 0 & 0 & 0 & 0 & 0 & 0 & 0 & 0 & 0 & 0 & 0 \\
 1 & 0 & 0 & 0 & 0 & \minus3 & 0 & 0 & 0 & 0 & 0 & 0 & 0 & 0 \\
 0 & 0 & 0 & 6 & \minus2 & 0 & 0 & 0 & 0 & 0 & 0 & 0 & 0 & 0 \\
 0 & 0 & 0 & 0 & 0 & 0 & 0 & 0 & 0 & 2 & 0 & 0 & 0 & 0 \\
 0 & 0 & 0 & 0 & 0 & 0 & 2 & 2 & 0 & 0 & 0 & 0 & 0 & 0 \\
 0 & 0 & 0 & 0 & 0 & 0 & 0 & 0 & 0 & \minus4 & 0 & 0 & 0 & 0 \\
 0 & 1 & \minus2 & 8 & \minus2 & 0 & 0 & 0 & 2 & 0 & 4 & 2 & 4 & 0 \\
\end{array}
\right],\\
\scriptsize
\arraycolsep=1pt\def\arraystretch{0.3}
\tilde{S}_2=\left[
\begin{array}{cccccccccccccc}
 \minus2 & 0 & 0 & 0 & 0 & 0 & 0 & 0 & 0 & 0 & 0 & 0 & 0 & 0 \\
 0 & 0 & 0 & 0 & 0 & 0 & 0 & 0 & 0 & 0 & 0 & 0 & 0 & 0 \\
 0 & 0 & \minus4 & 0 & 0 & 0 & 0 & 0 & 0 & 0 & 0 & 0 & 0 & 0 \\
 0 & 0 & 0 & 2 & 0 & 0 & 0 & 0 & 0 & 0 & 0 & 0 & 0 & 0 \\
 0 & 0 & 0 & 0 & \minus2 & 0 & 0 & 0 & 0 & 0 & 0 & 0 & 0 & 0 \\
 1 & 0 & 0 & 0 & 0 & \minus4 & 0 & 0 & 0 & 0 & 0 & 0 & 0 & 0 \\
 0 & 0 & 0 & 0 & 0 & 0 & \minus2 & 0 & 0 & 0 & 0 & 0 & 0 & 0 \\
 0 & 0 & 0 & 0 & 0 & 0 & 0 & \minus2 & 0 & 0 & 0 & 0 & 0 & 0 \\
 0 & 0 & 0 & 6 & 2 & 0 & 0 & 0 & \minus2 & 0 & 0 & 0 & 0 & 0 \\
 0 & 0 & 0 & 0 & 0 & 0 & 0 & 0 & 0 & \minus2 & 0 & 0 & 0 & 0 \\
 0 & 0 & 0 & \minus4 & 0 & 0 & 0 & 0 & 0 & 0 & 0 & 0 & 0 & 0 \\
 0 & 0 & 0 & 0 & 0 & 0 & 0 & 0 & 0 & 0 & 0 & 0 & 0 & 0 \\
 0 & 2 & 0 & 6 & 0 & 0 & 0 & 0 & 2 & 0 & 0 & 0 & 0 & 0 \\
 0 & 0 & 0 & 0 & 0 & 0 & 0 & 0 & 0 & 0 & 0 & 0 & 0 & \minus4 \\
\end{array}
\right],\
\tilde{S}_3=\left[
\begin{array}{cccccccccccccc}
 0 & 0 & 0 & 0 & 0 & 0 & 0 & 0 & 0 & 0 & 0 & 0 & 0 & 0 \\
 0 & 0 & 0 & 0 & 0 & 0 & 0 & 0 & 0 & 0 & 0 & 0 & 0 & 0 \\
 0 & 0 & 0 & 0 & 0 & 0 & 0 & 0 & 0 & 0 & 0 & 0 & 0 & 0 \\
 0 & 0 & 0 & 0 & 0 & 0 & 0 & 0 & 0 & 0 & 0 & 0 & 0 & 0 \\
 0 & 0 & 0 & 0 & 0 & 0 & 0 & 0 & 0 & 0 & 0 & 0 & 0 & 0 \\
 0 & 0 & 0 & 0 & 0 & 0 & 0 & 0 & 0 & 0 & 0 & 0 & 0 & 0 \\
 0 & 0 & 0 & 0 & 0 & 0 & 0 & 0 & 0 & 0 & 0 & 0 & 0 & 0 \\
 0 & 0 & 0 & 0 & 0 & 0 & 0 & 0 & 0 & 0 & 0 & 0 & 0 & 0 \\
 0 & 2 & 0 & 6 & 0 & 0 & 0 & 0 & 2 & 0 & 0 & 0 & 0 & 0 \\
 0 & 0 & 0 & 0 & 0 & 0 & 0 & 0 & 0 & 0 & 0 & 0 & 0 & 0 \\
 0 & 0 & 0 & 0 & 0 & 0 & 0 & 0 & 0 & 0 & 0 & 0 & 0 & 0 \\
 0 & 0 & 0 & 0 & 0 & 0 & 0 & 0 & 0 & 0 & 0 & 0 & 0 & 0 \\
 0 & 2 & 0 & 6 & 0 & 0 & 0 & 0 & 2 & 0 & 0 & 0 & 0 & 0 \\
 0 & 0 & 0 & 0 & 0 & 0 & 0 & 0 & 0 & 0 & 0 & 0 & 0 & 0 \\
\end{array}
\right]
\end{gather}
%\todo{add about boundary conditions}
We fix  the boundary conditions by considering the asymptotic coefficients at $s\to 4$. Most of the nontrivial boundary constants correspond to the naive values of the integrals at the threshold. The only exception is the leading threshold asymptotics of $j_3$, proportional to $\beta^{1-2\e}$. Namely, we explicitly calculate the following constants:
\begin{gather*}
    \tilde{j}_1^{\mathrm{th}} =1,\quad
    \tilde{j}_3^{\mathrm{th}} \sim \beta^{1-2\epsilon}\frac{\sqrt{\pi } \sin (\pi  \epsilon ) \Gamma \left(\epsilon -\frac{1}{2}\right)}{2 \Gamma (\epsilon -1)},\quad
    \tilde{j}_4^{\mathrm{th}}= \frac{\pi ^{3/2} \csc (\pi  \epsilon )}{2 \Gamma \left(\frac{3}{2}-\epsilon \right) \Gamma (\epsilon -1)},\quad 
    \tilde{j}_7^{\mathrm{th}}=\frac{-1+\epsilon}{2 (1-2 \epsilon)},\\
    \tilde{j}_8^{\mathrm{th}}=\tilde{j}_{10}^{\mathrm{th}}=\frac{1}{4} (1-\epsilon) \left[\psi\left({1}/{2}-\epsilon \right)+\gamma +2\ln 2\right]\,.
\end{gather*}
Here we again have chosen the overall factor so that $\tilde{j}_1^{\mathrm{th}} =1$. These boundary conditions, are sufficient to fix the specific solutions for the integrals $\tilde{\boldsymbol J}$. Using these solutions, we obtain for the ``bare'' cross section
%%TODO insert bare cross section.
\begin{multline}
    \left[\sigma_{\eegg}\right]_{\alpha^3,\,\text{bare}}=\\
    %2g/CrossSectionBare1: <-- this is needed for automatic check
    \frac{\alpha ^3}{s\beta}
     \Bigg\{
     \frac{1}{\epsilon}\left[\frac{2 (s-2) \left(s^2+4 s-8\right) {\ln^2 z}}{(s-4) s^2}+\frac{\left(s^3-6 s^2+16 s-48\right) {\ln z}}{s^3 \beta }-\frac{4  \left(s^2+4 s-6\right)}{s^2}\right]\\
     -\pi ^2 \frac{1+\beta^2}{\beta}\left[ 2-\beta^2+\frac{3-\beta^4}{2\beta}{\ln z}\right]
     +\S\Bigg[
     \frac{6\left(s^2+s-3\right)}{s^2\beta} \Re\bigg[
        8 \text{Li}_3\left(\tfrac{e^{\frac{i \pi }{3}}}{z+1}\right)
        +4\text{Li}_2\left(\tfrac{e^{\frac{i \pi }{3}}}{z+1}\right) {\ln (s\,z)}
        \\
        -\text{Li}_2\left(-z\right) {\ln s}
        - \frac12 {\ln \frac{(s-1)^2}{s}}\, {\ln s}\ {\ln z}
        -\frac1{9}{\ln^3 z}
     \bigg]
     -\frac{2 \left(2 s^2-s-9\right) }{s^2\beta}\bigg[
        4\text{Li}_3\left(\tfrac{1}{z+1}\right)
        +\text{Li}_2(-z) {\ln s}
        \\ 
        +\frac14 {\ln^2 s}\, {\ln z}
        +\frac7{12}{\ln^3 z}
     \bigg]
     +\frac{(s-3) (2 s+5)}{3 s^2\beta}\left[{\ln^2 z}-\pi ^2 \right]{\ln z}
     -\frac{2 \left(3 s^3+2 s^2-40 s+32\right)}{(s-4) s^2}{\ln s}\, {\ln^2 z}\\
     -\frac{2 (s+2)}{s} \text{Li}_2(-z) {\ln z}
     +\frac{2\left(3 s^3+14 s^2-48 s+48\right)}{s^3\beta} \text{Li}_2(1-z)
     +\frac{20}{s\beta} \text{Li}_2(-z)\\
     -\frac{7 s^2+2 s-56}{2 (s-4) s} {\ln^2 z}
     -\frac{2 s^5-25 s^3-12 s^2+80 s-48}{(s-1)^2 s^3\beta} {\ln s} {\ln z}
     +\frac{s+2}{2 s}\pi ^2 \\
     +\frac{5 s^3+2 s^2-32 s+24}{(s-1) s^2} {\ln s}
     +\frac{10 s^4-35 s^3+86 s^2-128 s+64}{(s-1) s^3\beta} {\ln z}
     -\frac{ (s+8) (3 s-4)}{s^2}
     \Bigg]
     \Bigg\}
    %/2g/CrossSectionBare1: <-- this is needed for automatic check
    \,.
\end{multline}

The onshell renormalization procedure is described in the literature, see, e.g., Ref. \cite{Grozin:2005yg}. For our setup, this means that the cross section expressed in terms of the physical parameters reads
\begin{equation}
    \sigma_{\eegg}=\sigma_0+\left[\sigma_{\eegg}\right]_{\alpha^3,\,\text{bare}}
    +(Z_{\psi}^2Z_A^2Z_{\alpha}^2-1)\sigma_0
    +\delta \sigma_{m},
\end{equation}
where $Z_{\psi}$, $Z_A$, and $Z_{\alpha}$ are the onshell renormalization constants for the electron field, photon field, and coupling constant, respectively. Since  $Z_AZ_{\alpha}=1$ due to Ward identity, we have $Z_{\psi}^2Z_A^2Z_{\alpha}^2-1\approx 2\delta Z_{\psi}=-2\frac{(4 \pi  \alpha ) (3-2 \epsilon ) \Gamma (\epsilon )}{(4 \pi )^{2-\epsilon } (1-2 \epsilon )}\left(\frac{e^{\gamma_E}}{4 \pi }\right)^{\epsilon }$. Note that an additional factor $\left(\frac{e^{\gamma_E}}{4 \pi }\right)^{\epsilon }$ as compared to Ref. \cite{Grozin:2005yg}, corresponds to the chosen loop measure. The term $\delta \sigma_{m}$ corresponds to the contribution of diagrams $(v), (x)$ in Fig. \ref{fig:amplitude2} associated with the mass counterterm. On those diagrams the cross corresponds to the vertex $i\delta m=i \frac{(4 \pi  \alpha ) (3-2 \epsilon ) \Gamma (\epsilon )}{(4 \pi )^{2-\epsilon } (1-2 \epsilon )}\left(\frac{e^{\gamma_E}}{4 \pi }\right)^{\epsilon }$. We have
\begin{multline}\label{eq:mct}
\delta\sigma_m=
    %2g/MassCounterterm: <-- this is needed for automatic check
\frac{\alpha ^3}{s\beta} \S\bigg\{
    \frac{3}{\epsilon}\left[
        \frac{1}{s\beta}\left(1+\beta ^4\right) {\ln z}
        +\frac{1}{2}\left(3-\beta^4\right)
   \right]
   -\frac{3}{s\beta}\left(1+\beta ^4\right)\left[
        2\text{Li}_2(1-z)+ {\ln s}\, {\ln z}
   \right]
   \\
   -\frac{3}{2} \left(3-\beta^4\right) {\ln s}
   -\frac{1}{2\beta} \left(7-\beta^2-2 \beta ^4+2 \beta ^6\right){\ln z}
   -2 \beta ^4
   -\frac{12}{s}
   \bigg\}
    %/2g/MassCounterterm: <-- this is needed for automatic check
    \,,
\end{multline} 
where we have neglected terms suppressed in $\e$.
Note that the renormalized cross section $\sigma_{\eegg}$ still contains $\e^{-1}$ terms due to infrared divergence. In order to obtain the observable cross section $\sigma_{\eegg}^{\mathrm{f}}(\omega_0)$ we have to add the soft-photon contribution $W^{\mathrm{f}}(\omega_0)\sigma_0$, where $W^{\mathrm{f}}(\omega_0)$ is defined in Eq. \eqref{eq:Wf} for $\mathrm{f}=\mathrm{cmf},\mathrm{rf}$. Finally, we obtain Eqs. \eqref{eq:CrossSection2cm} and \eqref{eq:CrossSection2r} for the cross sections $\sigma_{\eegg}^{\mathrm{cmf}}(\omega_0)$ and $\sigma_{\eegg}^{\mathrm{rf}}(\omega_0)$, respectively.

\section{Conclusion}
%TODO: remove commented lines in the table
\begin{table}[t]
    \centering\begin{tabular}{|l|l|r|r||r|r|}
        \hline
        \multirow{2}{*}{$\displaystyle{\sqrt{s}\over m_e}$} & \multirow{2}{*}{$\displaystyle{\omega_0 \over m_e}$} & \multicolumn{2}{c||}{$\sigma_{\eeggg}^{\mathrm{cmf}}$(mb)} &  \multicolumn{2}{c|}{$\sigma_{\eeggg}^{\mathrm{rf}}$(mb)}  \\ \cline{3-6}
        & & Exact, Eq.\eqref{eq:CrossSection3cm} & Monte-Carlo & Exact, Eq.\eqref{eq:CrossSection3r} & Monte-Carlo \\     	
        \hline
        $2.001$                 & $0.01$ &   $10.682$ &   $10.672(31)$ &         $10.682$ &         $10.671(32)$ \\
        $2.01$                  & $0.01$ &    $3.551$ &    $3.551(11)$ &          $3.551$ &          $3.551(11)$ \\
        $2.1$                   & $0.01$ &   $1.8194$ &   $1.8222(57)$ &         $1.8297$ &         $1.8321(58)$ \\
        $5$                     & $0.01$ &   $3.0238$ &   $3.0229(96)$ &          $3.430$ &          $3.428(11)$ \\
        $10$                    & $0.01$ &    $1.843$ &    $1.847(18)$ &         $2.2078$ &          $2.208(10)$ \\
        $50$                    & $0.01$ &   $0.3158$ &   $0.3128(31)$ &         $0.4045$ &         $0.4037(41)$ \\
        %        $50$                    & $0.03$ &   $0.??$ &   $0.??$         &         $0.3570$ &        $0.3544(35)$      \\
        $50$                    & $0.1$ &   $0.2161$ &   $0.2151(21)$         &         $0.3048$ &        $0.3023(30)$      \\
        $50$                    & $1$    &   $0.1163$ &   $0.1157(11)$ & \uwave{$0.2051$} & \uwave{$0.1854(19)$} \\
        $50$                    & $5$    &  $0.04663$ &  $0.04698(47)$ & \uwave{$0.13539$}& \uwave{$0.09783(98)$} \\
        %        $100$                   & $0.01$ &  $       $ &  $           $ & $0.1628$         & $0.14852(15)$\\
        $100$                   & $0.1$ &  $0.08897$  &  $0.08842(88)$ & $0.1268$         & $0.1289(13)$\\
        $100$                   & $1$ &  $0.05296$  &  $0.05181(52)$   & \uwave{$0.09084$}& \uwave{$0.08371(84)$}\\
        $100$                   & $10$ &  $0.01695$  &  $0.01682(17)$   & \uwave{$0.054833$}& \uwave{$0.03878(39)$}\\
        %    	$200$                   & $0.01$ &  $0.04639$ &  $0.04640(46)$ &       $0.061269$ & $0.05248(52) $ \\ 
        %    	$200$                   & $1$    &   $0.0218$ &    $0.0214(2)$ &                  &                      \\
        %        $200$                   & $20$   & $0.005821$ & $0.005807(29)$ &     $0.02069$    &$0.01417(14)$    \\
        %24 min 
        \hline 
    \end{tabular}
    \caption{Comparison of our analytic result, Eq. \eqref{eq:CrossSection3cm}, for $\sigma_{\eeggg}^{\mathrm{cmf}}$ with that of Monte-Carlo integration performed using \texttt{Cuba} library \cite{hahn2016concurrent}. The errors in last column are those provided by \texttt{Cuba}.}
    \label{tab:comparison}
\end{table}

In the present paper we have calculated the total cross sections $\sigma_{\eeggg}^{\mathrm{f}}(\omega_0)$ and $\sigma_{\eegg}^{\mathrm{f}}(\omega_0)$ for arbitrary energies with $O(\alpha^3)$ accuracy. The energy cut $\omega_0$ for soft photons has been applied in the center-of-mass frame ($\mathrm{f=cmf}$) and in the rest frame of the electron ($\mathrm{f=rf}$). We have found errors in the high-energy results available in the literature for $\sigma_{\eeggg}^{\mathrm{cmf}}(\omega_0)$.

As an additional check of our results for the 3-photon annihilation cross section, we have performed a numerical integration of the differential  cross section using the \texttt{Cuba} library \cite{hahn2016concurrent}. Table \ref{tab:comparison} demonstrates a perfect agreement of  our results with the numerical calculation. The exception are the points where the minimal photon energy $\omega_0$ is of the order of electron mass in the rest frame of the initial electron (marked with wavy lines in the table). This is, of course, quite expected as the photon with energy of the order of electron mass in electron rest frame can have energy of the order of $\sqrt{s}$ when one passes to the center-of-mass frame (thus, the soft-photon approximation breaks).

\paragraph*{Acknowledgments} I am grateful to V. Fadin and A. Milstein for the interest to this work and stimulating discussions, and to A. Grozin for clarifying some issues related to onshell renormalization scheme. I acknowledge the support from the ``Basis'' foundation for theoretical physics and mathematics and from the Russian Science Foundation (grant 20-12-00205) for the calculations related to $\eeggg$ and to $\eegg$, respectively.

\appendix

\section*{Ultrarelativistic limit from approximate differential cross section}\label{sec:UR}

Ref. \cite{berends1981distributions} (as well as Ref. \cite{eidelman1978e+}) used the approximate differential cross section 
\begin{equation}
    d\sigma_{\eeggg}^{\mathrm{cmf}}=\frac{(4\pi\alpha) ^3}{s}s\overline{\left|M\right|^2} \frac{d\Phi_{3\gamma}}{3!\,2s},
\end{equation} 
where
\begin{equation}
    d\Phi_{3\gamma}=(2\pi)^4\delta^{(4)}(p_++p_--k_1-k_2-k_3)\prod_{i=1}^3\frac{d\boldsymbol{k}_i}{(2\pi)^32\omega_i}\,,
\end{equation}
is the phase space of the final particles, and 
\begin{equation}
    s\overline{\left|M\right|^2}\approx 8\frac{\kappa_{1-}^2+\kappa_{1+}^2}{\kappa_{2-}\kappa_{3-}\kappa_{2+}\kappa_{3+}}
    -\frac{8}{\gamma^2}\left[\frac{\kappa_{3+}}{\kappa_{1-}^2\kappa_{2+}}+\frac{\kappa_{3-}}{\kappa_{1+}^2\kappa_{2-}}\right]+\text{permutations}\,.
\end{equation} 
Here $\kappa_{i\pm}= 4k_i\cdotp p_{\pm}/s=\frac{\omega_i}{\varepsilon_-}(1\pm \beta \cos\theta_i)$. At large angles and $\omega_i\sim \varepsilon_-$ we have $\kappa_{i\pm}\sim 1$, while for $\theta_i\lesssim \gamma^{-1}$ ($\pi-\theta_i\lesssim \gamma^{-1}$) we have $\kappa_{i-}\sim \gamma^{-2}$ ($\kappa_{i+}\sim \gamma^{-2}$). Thus the second term, formally suppressed by $\gamma^{-2}=4/s$, contributes in the region when the momentum of one of the final photons is close to forward or backward direction, when one of the squared denominators in square brackets gives an amplifying factor $\sim \gamma^{4}$. Thus, we have the following power counting: $\frac{d^2\theta_1}{\gamma^2\kappa_{1-}^2}\sim \frac{\gamma^{-2}}{\gamma^2(\gamma^{-2})^2}\sim 1$.
We have checked that the integration of this expression for $s\overline{\left|M\right|^2}$ indeed leads to the result \eqref{eq:berends} of Ref. \cite{berends1981distributions}\footnote{Therefore, Ref. \cite{eidelman1978e+} also contains a technical mistake.}. So, the origin of the discrepancy of our result with that of Ref. \cite{berends1981distributions} can be only in the initial expression for the differential cross section. Indeed, a thorough inspection of the exact expression for the differential cross section from Ref. \cite{mandl1952theory} has revealed the overlooked in Refs. \cite{berends1981distributions,eidelman1978e+} terms which contribute to the total cross section. Namely, in $s\overline{\left|M\right|^2}$ one has to take into account also the terms 
\begin{equation}
    -\frac{8}{\gamma^2}\left[\frac{1}{\kappa_{1+}\kappa_{2-}\kappa_{3-}}+\frac{1}{\kappa_{1-}\kappa_{2+}\kappa_{3+}}\right]+\text{permutations}
\end{equation}
These terms contribute in the kinematic region where simultaneously two photons have small scattering angles. Then the third photon necessarily has scattering angle close to $\pi$ and we have the following power counting: $\frac{d^2\theta_2d^2\theta_3}{\gamma^2\kappa_{1+}\kappa_{2-}\kappa_{3-}}\sim \frac{\gamma^{-2}\gamma^{-2}}{\gamma^2\gamma^{-2}\gamma^{-2}\gamma^{-2}}\sim 1$. We have checked that these terms, overlooked in Refs. \cite{eidelman1978e+,berends1981distributions}, give exactly the contribution $-\frac{4\alpha^3 \pi^2}{3s}$ to the total cross section, in agreement with our asymptotics \eqref{eq:3ghigh}.

\bibliographystyle{Refs/JHEP}
\providecommand{\href}[2]{#2}\begingroup\raggedright\endgroup
\end{document}